\documentclass[prl,twocolumn,aps,showpacs]{revtex4} 

\usepackage{graphicx}
\usepackage{dcolumn}
\usepackage{bm}

\begin{document}

\title{To how many politicians should government be left?}

\author{Peter Klimek$^{1}$, Rudolf Hanel$^{1}$, Stefan Thurner$^{1,2}$}
\email{thurner@univie.ac.at} 

\affiliation{
$^1$ Complex Systems Research Group; HNO; Medical University of Vienna; 
W\"ahringer G\"urtel 18-20; A-1090; Austria \\
$^2$ Santa Fe Institute; 1399 Hyde Park Road; Santa Fe; NM 87501; USA\\
} 


\begin{abstract}
The quality of governance of institutions, corporations and countries
depends on the ability of efficient decision making within the
respective boards or cabinets. Opinion formation processes within
groups are size dependent. It is often argued - as now e.g. in the
discussion of the future size of the European Commission - that
decision making bodies of a size beyond 20 become strongly
inefficient. We report empirical evidence that the performance of
national governments declines with increasing membership and
undergoes a qualitative change in behavior at a particular group
size. We use recent UNDP, World Bank and CIA data on overall government efficacy,
i.e. stability, the quality of policy formulation as well as human
development indices of individual countries and relate it to the
country's cabinet size. We are able to understand our findings
through a simple physical model of opinion dynamics in
groups.
\end{abstract}

\maketitle

\section{Introduction}
Honorable statesmen, like Charles de Gaulle or Chester Bowles, arrived at the conclusion that 
'politics is too important to be left to politicians'. 
The highest executive power in today's political landscape is mostly conferred upon committees called 
cabinets -- the countries' governments -- consisting of people having, according to Robert Louis Stevenson, 
the only profession for which no preparation is thought necessary. 
It is natural to ask to how many of them government can be left without furnishing a democratic collapse. 
The question to how many individuals government should be left to ensure democratic effectiveness 
was first tackled in a semi-humorous attempt by the British historian C. Northcote Parkinson \cite{Park57}. 
His investigations lead to what is now known as the 'Coefficient of Inefficiency', conjecturing that a 
cabinet loses political grip, due to an inability of efficient 
decision-making, as soon as its membership passes a critical size of 19-22. 
This result's validity applies to decision-making in groups in general. 

We show that Parkinson's conjectures about cabinet sizes and government efficiency hold empirically to remarkable 
levels of significance. By relating cabinet size to several governance indicators, assembled by the UNDP (the Human Development Indicator\cite{hdr}), 
the CIA \cite{chiefs} and the World Bank \cite{Kaufmann07},
we confirm the hypothesis that the higher number of members in the highest executive committee,  
countries are more likely to be political less stable, less efficient and less developed. 
We are able to understand the origin of the critical size of 20 members in a simple opinion 
formation model \cite{Klimek07}  extended to small-world network interactions in small groups.

\begin{figure*}[tb]
 \begin{center}
 \includegraphics[height=58mm]{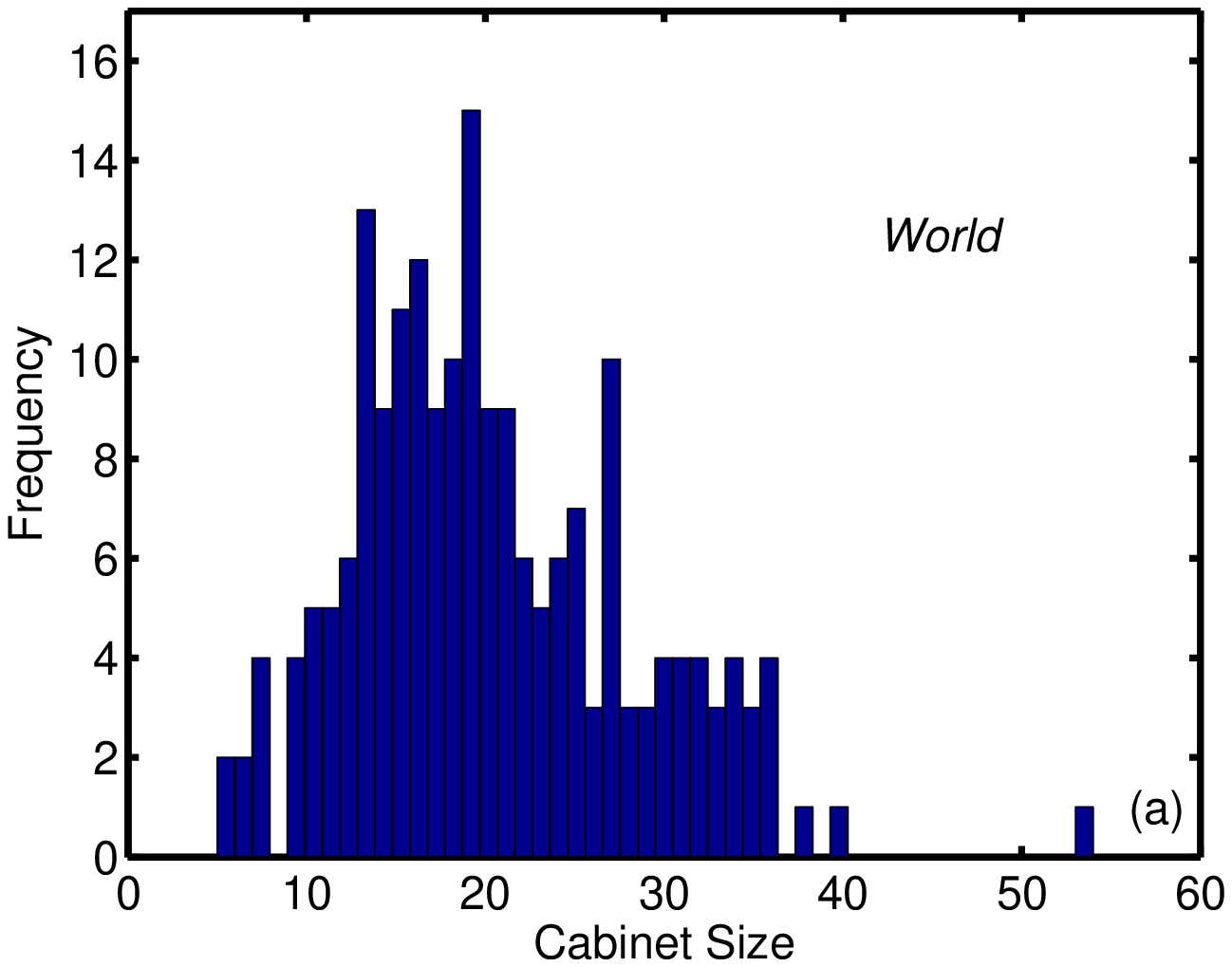}
 \includegraphics[height=58mm]{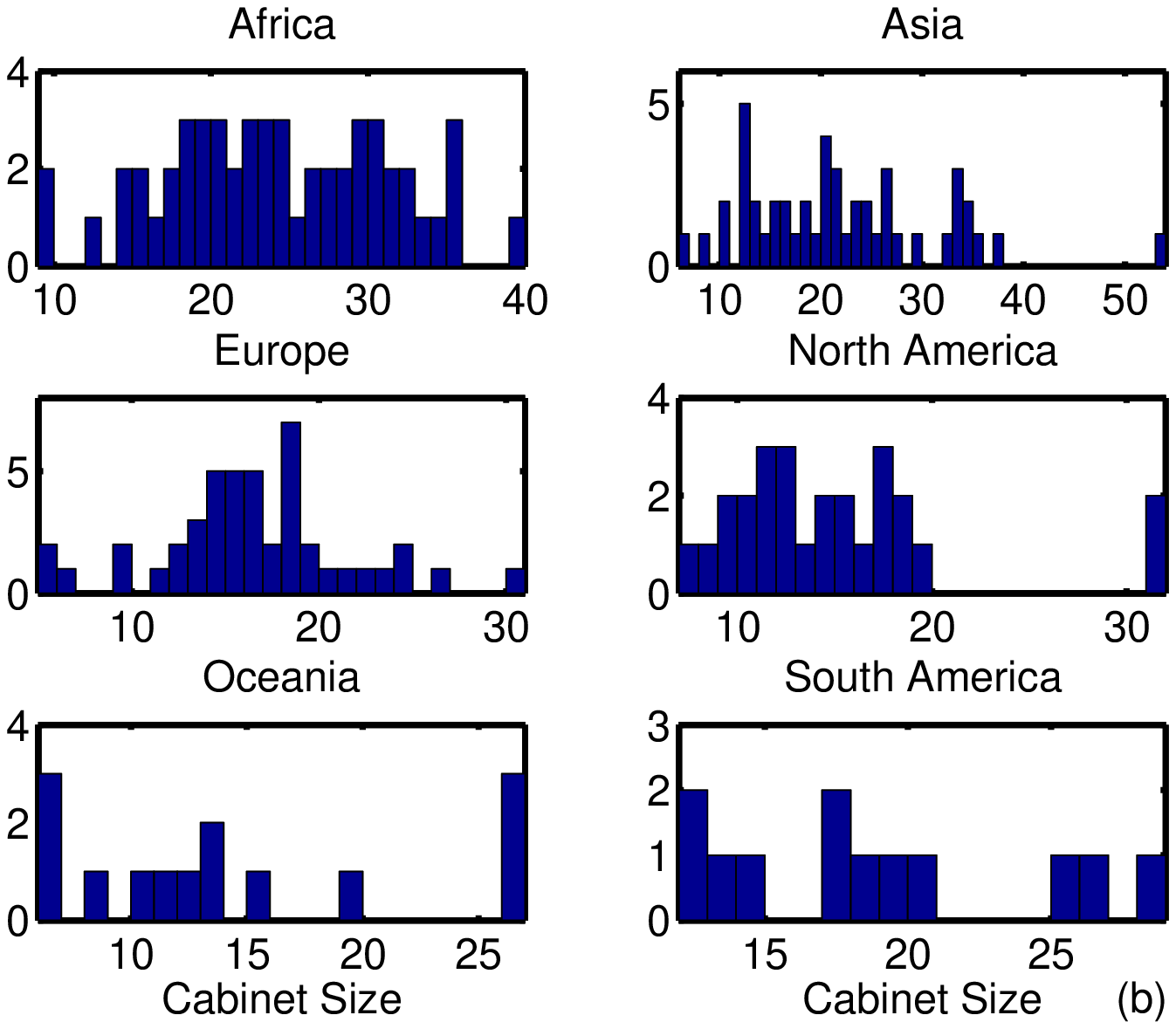}    
  \end{center}
 \caption{Histograms of (a) the world's cabinet sizes, (b) for each continent show that the cabinet sizes for Europe, America and Oceania follow the same pattern with the vast majority of countries lying below 20, whereas in Africa and Asia cabinets tend to grow beyond this point. There is no cabinet with eight members.}
 \label{hist_ind}
\end{figure*}

\section{Cabinet sizes and efficiency: empirical findings}

We determine the actual number of members of the highest executive committee, the cabinet, for 197 self-governing countries and territories using data provided by the CIA \cite{chiefs}. For a complete listing
of them see table 1
 in the appendix. Cabinets vary between 5 and 54 members with a clearly visible peak between 13 and 20. All except three countries (Pakistan, Democratic Republic of Congo and Sri Lanka) are found in the range between 5 and 36. 
It is worth noting that all countries 
avoid cabinets with 8 members, a curious fact that was observed already some fifty years ago \cite{Park57}.

\begin{figure*}[tb]
 \begin{center}
\includegraphics[height=54.9mm]{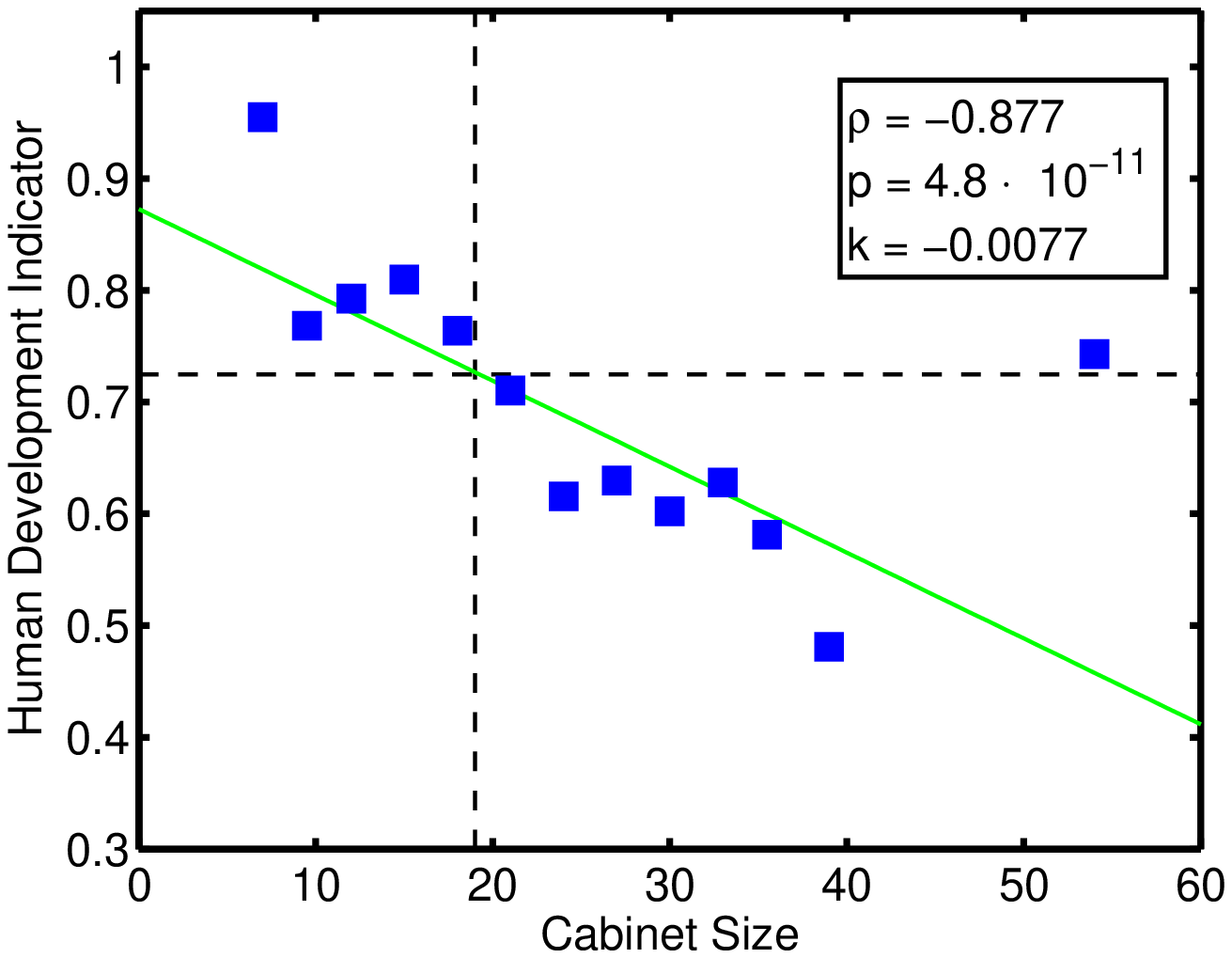}
 \includegraphics[height=54.9mm]{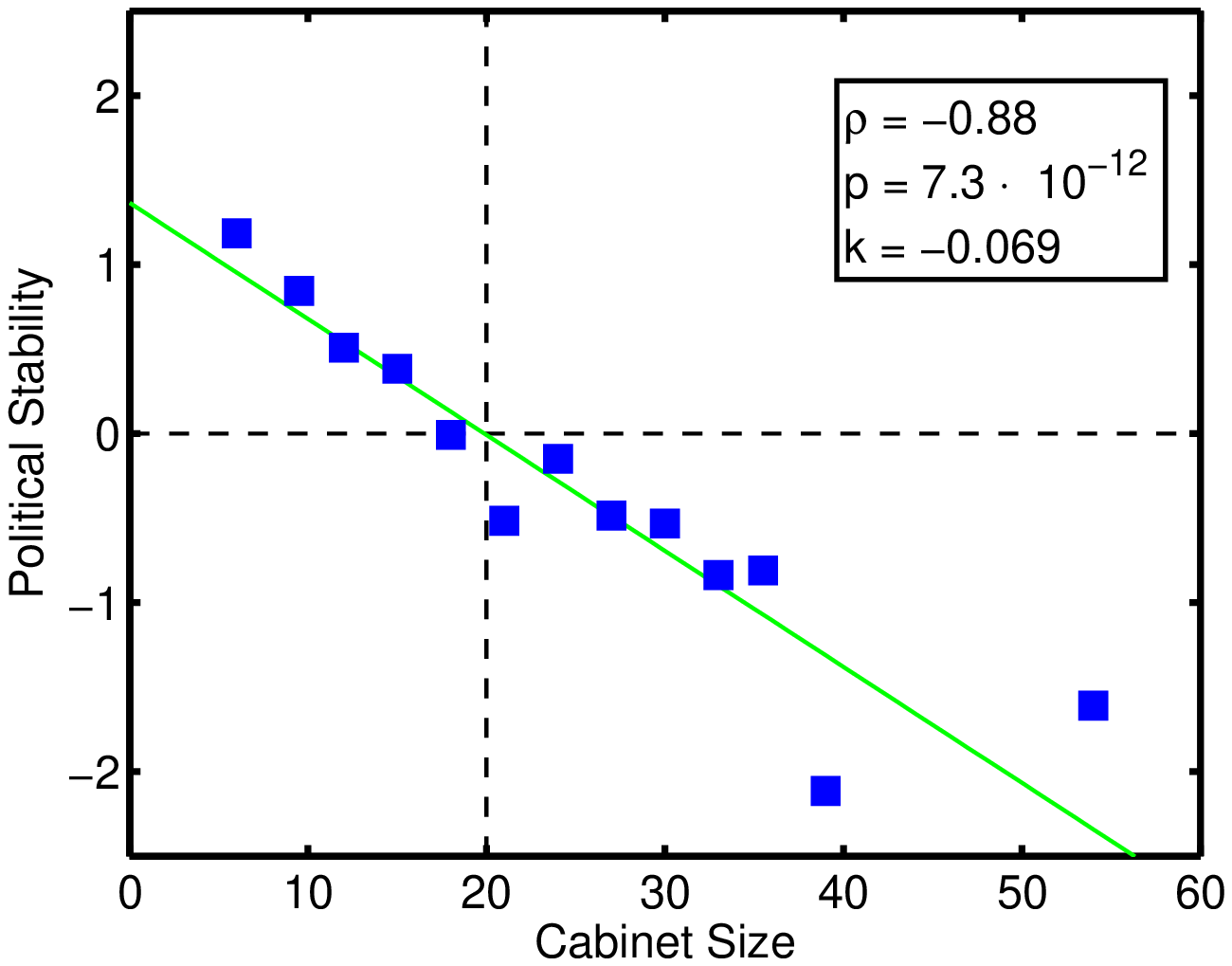}     \\
 \includegraphics[height=54.9mm]{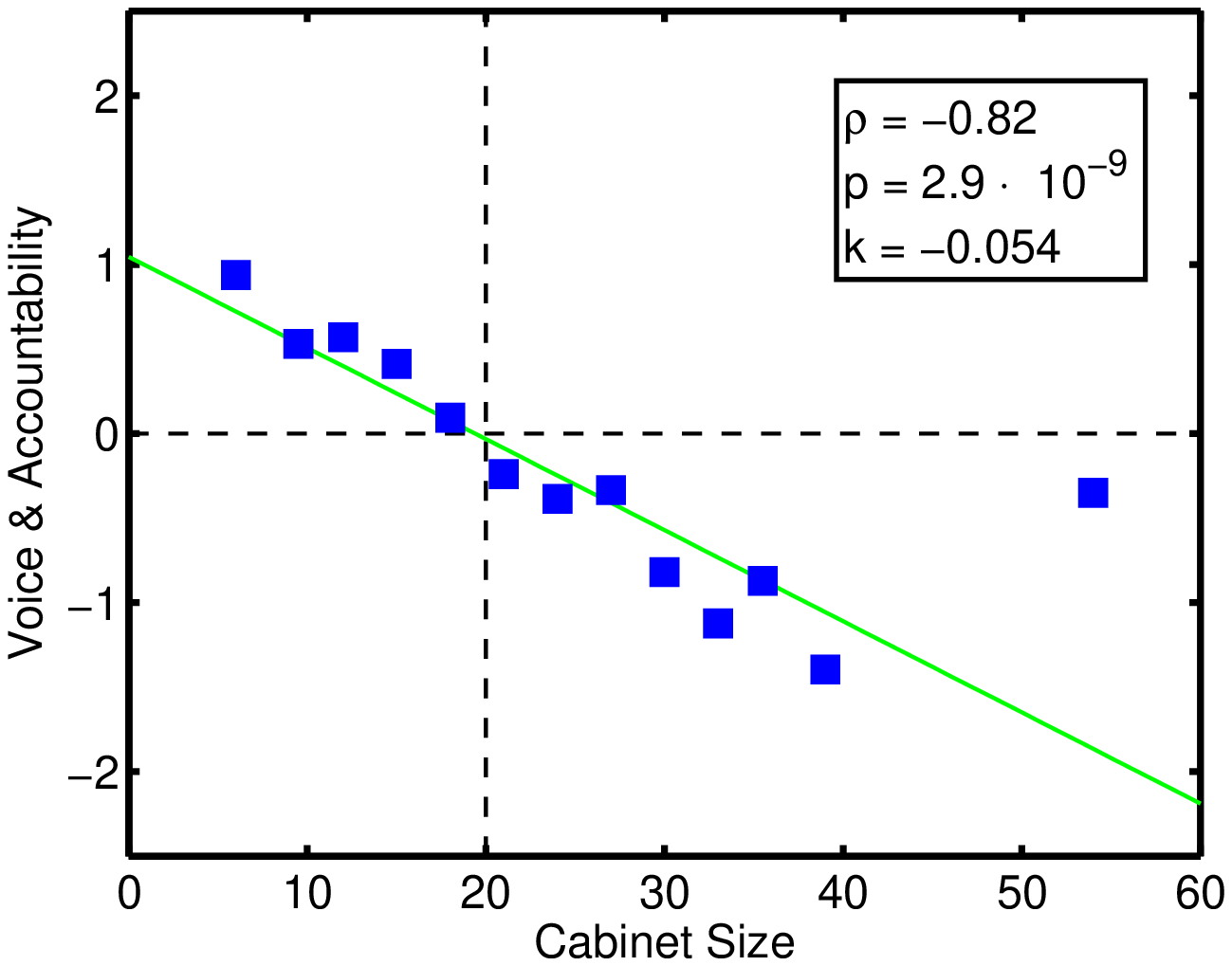}
 \includegraphics[height=54.9mm]{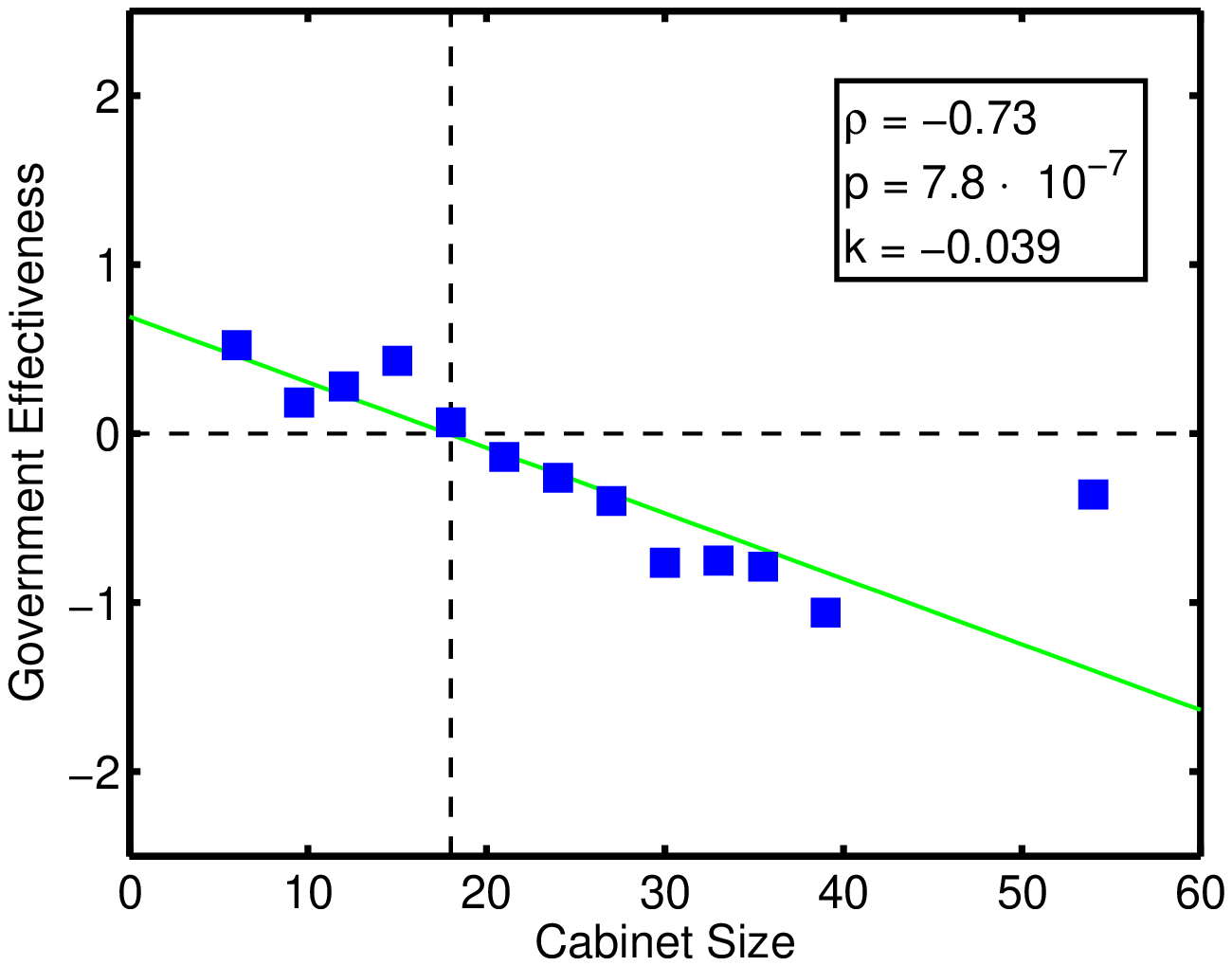} 
  \end{center}
 \caption{Cabinet size is negatively correlated with the Human Development Indicator, Political Stability, Voice \& Accountability and Government Effectiveness. For each indicator the line separating countries ranking above and below the global average lies around 20. The correlation is highly significant 
to a $p$-valiue of $p \leq 10^{-6}$.}
 \label{gi_pl}
\end{figure*}

To determine whether cabinet size can serve as an indicator for efficient policy making 
we compare it with indicators reflecting complex issues of states which -to get advanced reasonably- 
need a certain consensus within the political leadership.  
 One such indicator is the UNDP Human Delevopment Indicator\cite{hdr} (HDI) which assesses a country's 
achievement in different areas of human development. It is composed of the GDP, life expectancy at birth, 
the literacy and the gross enrolment ratio. 
A second indicator is assessed on behalf of the World Bank\cite{Kaufmann07}, measuring a country's 
governance along three dimensions: Political Stability (PS, indicating the likelihood that the government will be destabilised or overthrown), Voice \& Accountability (VA, quantifying to which extent citizens can select their government) and Government Effectiveness (GE, measuring the qualitiy of policy formulation and implementation). Note that none of these indicators includes any prior dependence on the cabinet size.  
 Figure \ref{gi_pl} shows the average values for these 4 indicators versus cabinet size. 
Note that the value for these indicators falls below the global average (line) when cabinet size exceeds 20
(Parkinson's coefficient of inefficiency).
Interestingly the frequency of cabinet-sizes peaks at this point and slightly below, see figures
\ref{hist_ind} (a),(b).
This indicates that cabinets are most commonly constituted with memberships close to Parkinson's coefficient, but not above it and thus lends further support to the conjecture that a cabinet's functioning undergoes a remarkable change at this point. 
These observations strongly suggest a correlation between increasing cabinet size and a declining
overall quality in governance and achievements for human development.
To assert statistical significance of the data we compute the correlation coefficient of size and 
the 4 indicators, and the $p$-value for the null-hypothesis that size and indicator are not correlated. 
For the HDI, PS, VA, and GE  we find correlation coefficients 
of $\rho=-0.88,-0.88,-0.82,-0.73$  
and significance levels  
of $p=4.8 \times 10^{-11}, 7.3 \times 10^{-12}, 2.9 \times 10^{-9}, 7.8 \times 10^{-7}$, respectively. 
Our results are thus significant against the null-hypothesis 
up to a $p$-value of $p \leq 10^{-6}$.
To exclude the possibility that we observe this due to a trivial super-correlation 
with e.g. size of the countries, we compute the corresponding 
correlation coefficient and $p$-value 
for the area ($\rho=0.24$, $p=0.16$) and population ($\rho=0.15$, $p=0.40$), i.e.  
no significant correlations.

\section{Opinion formation and group size}


How can these facts be understood? Why should a cabinet size around 20 be special in the sense that 
it separates countries ranking above and below the global average of the studied indicators? 
The idea of this paper is to show in a simple model that in opinion formation processes there exists a critical 
number of individuals, above which it becomes exceedingly difficult to reach consensus in the group. 

In general cabinets are subject to a law of growth. This has been elaborated in detail 
for British cabinets from the year 1257 up to the 20th century by Parkinson \cite{Park57}. 
In a sense cabinets reflect the most important interest groups in a country. 
Besides core ministries (like finance, inner and outer affairs, etc.), which exist in nearly all countries, 
some interests strongly depend on the region's characteristics. 
OPEC countries, for example, sustain a ministerial post for 
petroleum; countries with mixed ethnicities sometimes have a minister for each of 
them. A secretary for land mining or aviation will more probably be found in Africa than in 
Europe, to name only a few examples. Also the political climate is represented, e.g. 
the number of parties taking part in the government. On the one hand there is always pressure 
from outside groups seeking to be included and represented in decision-making processes, on 
the other hand it is obvious that the larger the decision-making body, the 
more difficult consensus is reached.
It is one of the classical challenges of governance to find a balance between these two competing forces:
wide representation and effective leadership.

To become more quantitative let us ask how the introduction of one additional member in a decision making body 
alters the body's ability to reach consensus upon executing a certain policy? 
If one additional voting member would lead to a significant decrease in consensus finding, 
there should be resistance to enlargement, if an additional member does not further complicate 
the opinion formation process, there should be no reason to exclude him/her. 

In case there exists a characteristic group size below which adding one member significantly decreases the ability to reach consensus, and above which this incremental decrease becomes smaller, it is reasonable to  conjecture that this 
characteristic size is critical for the functioning of a cabinet. Above this critical size there is less restriction to the admission of more representatives due to outside pressure, which in turn implies that a loss in efficiency is more likely. 
If a cabinet exceeds this point (coefficient of inefficiency) it gradually loses its ability 
to be an institution where decisions are reached and remains merely a nominal executive. 
In this case the effective executive power might not be in the hands of governments anymore.
We now show that such a critical point does exist within a large class of simple opinion formation models. 

\begin{figure}[tb]
 \begin{center}
 \includegraphics[height=65mm]{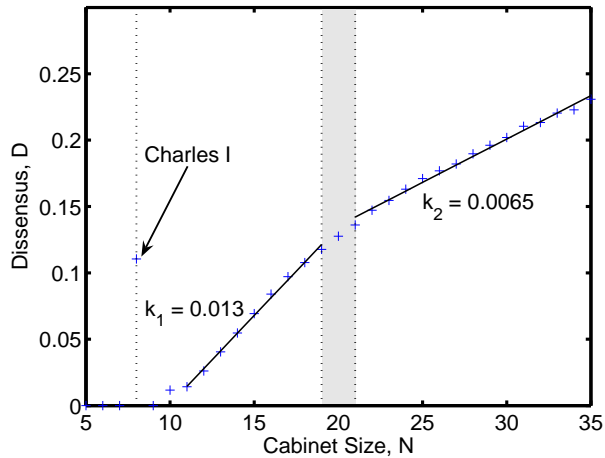}    
 \end{center}
 \caption{Simulation results for the dissensus parameter $D\left(N\right)$ versus cabinet size. For $N<10$
consensus is always reached except in the 'Charles I' scenario for $N=8$. 
As cabinet size increases dissensus becomes more likely. For $10\leq N<18$ 
each new member adds a dissensus-increment of $k_1=0.013$. 
The range between 19 and 21 (position of the conjectured coefficient of inefficiency) is shaded. 
Beyond this size the increment in dissensus by each new member is lowered to $k_2=0.0065$, which 
confirms the existence of a critical point. This  point separates two scenarios where in the first an increase in size has 
a comparably large negative impact on efficiency, an effect that diminishes in the second scenario 
where the admission of a new member 
has a minor effect.}
 \label{DSsimu}
\end{figure}

\section{A model for opinion formation in small-world groups}

In recent years physics has repeatedly crossed disciplinary boundaries toward a  quantitative understanding of 
social phenomena\cite{Helbing95,Schweitzer97,Castellano07}. A topic of mayor interest is to uncover 
the relevant mechanisms driving collective decision-making processes, the study of opinion formation 
models \cite{Liggett99}. The system is composed of interconnected agents, holding an internal state e.g. a 
binary opinion (like a spin in the Ising model), which interact by a given microscopic dynamical rule 
\cite{Galam86, BenNaim96,Axelrod97,Krapivsky03,Castellano05}. These local rules quantify 
the social influence individuals have upon each other. 
Depending on how these rules and inter-agent networks are specified, the system will evolve either 
toward a state given by maximal consensus \cite{Lambiotte07}, or alternatively the system may get 
stuck in a so-called 'frozen state' which is usually strongly determined by the initial conditions of the 
system\cite{Klimek07}.    
In the latter model the group is composed of $N$ individuals (nodes in a network), 
each one holding an internal state 0 or 1, for example a binary (yes/no) vote on a given topic.
Two agents who have social of informational influence upon each other are connected by a link in the network. 
For the inter-agent network we chose a small-world network\cite{Watts98}  where each node can potentially influence 
$k$ other nodes in its local neighborhood and, with some probability $L$, also nodes in the more distant 
neighborhood. For example, imagine agents having the same party affiliation (local neighborhood)
where they can influence each other in debates etc. With a certain probability ($L$) these agents 
might also talk to cabinet members of the opposite party, due to  e.g. overlapping responsibilities, sympathy, etc. 
An impressive number of social networks was shown to be of the small-world type 
\cite{Watts99},
for the remainder we consider this network as static over time. 
As a dynamical rule we implement a 'majority rule' with a predefined threshold \cite{Klimek07,Watts02}  $h \in \left(0.5, 1 \right]$. Here one node adopts the state of the majority of its neighbours only if this majority exceeds the fraction of its neighbors $hk$, otherwise the node's internal state stays unchanged. 
The threshold $h$ takes statistically account of various determinants whether an agent will conform to the majority's opinion. These determinants include the social status or prestige of the neighbors, the importance of 
the decision or the prepotency of the group's induced response\cite{Asch52, Kelman58}.
For $h>0.5$ the pure majority rule is recovered\cite{Galam86}. 
We choose $L=0.1$, $h=0.6$ and $k=\min\left[N-1,8\right]$. 
Parameter dependence of our results is discussed in the appendix, however, most findings are robust.

The evolution of this model is given by a random sequential application of the dynamical rule. 
In one iteration the described update procedure is applied once to each node in a random order. 
After a sufficient number of iterations the system will reach a stationary state where no more updates take place. 
The question here is whether this state is consensus, i.e. all nodes are in the same internal state, or not. 
The initial condition is determined by the fraction of nodes in the two respective states, 
let us call the number of nodes initially in 0 $A_i$ and the final population in this state $A_f$. 
For our purposes we want to determine the group's general ability to avoid dissensus.
Therefore we define (as the order parameter) the 'dissensus' parameter, 
\begin{equation}
D\left( N \right) = \left \langle \Theta \left(1- \frac{\max\left(A_f,N-A_f\right)}{N}\right) \right \rangle_{A_i}
\quad ,
\label{Dn}
\end{equation}
where $\Theta \left(x\right)$ is the Heaviside step function and $\langle \cdot \rangle_{A_i}$ denotes the average over all possible initial conditions. $A_i$ is drawn with uniform probability from $\left(0,1,\dots N \right)$. According to this the opinions are randomly assigned to the individual nodes. 
$D\left(N\right)$ is the expectation value of a final state without consensus and measures the group's proneness to end up in dispute. It only depends on the group-size $N$. 
Dissensus vs. $N$ is shown in figure \ref{DSsimu} for fixed $k$. 
For groups of less than 10 members consensus can always be reached, with the notable exception of $N=8$ (we refer to this case as 'Charles I'. Why? See explanation below). For $10<N<20$ increase of group size leads to increasing dissensus with a constant rate (slope) of $k_1=0.013$. This behaviour changes at $N\sim 20$, where increments become 
considerably smaller; a linear fit yields a slope of $k_2=0.0065$.

\begin{figure}[tb]
 \includegraphics[height=42mm]{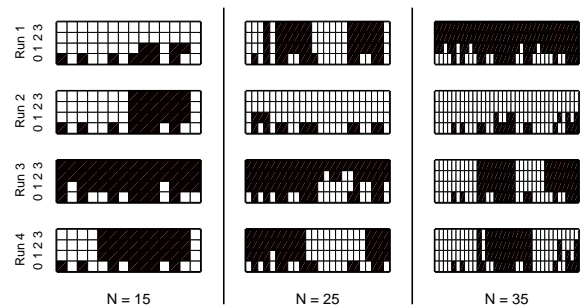}    
 \caption{
Evolution of the opinion formation process in groups of three different sizes ($N=15,25,35$). 
Each individual is a box, whose opinion is either black or white.
For each value of $N$ we show 4 independent update-runs, all starting with the same initial configurations 
and the same network. In each run we show timesteps 0,1,2,3, where time 0 corresponds to the initial
configuration. Line 1 is obtained by the iterative application of the opinion formation protocol on the 
initial configuration in a random sequence of updates.
The next lines (2,3) are obtained in the same way. The particular 
sequence is seen to play a crucial role for the final state at time 3. 
Two different trajectories leading to consensus (all colors white or black at time 3) 
or dissensus (mixed colors) are shown for each size. It becomes apparent that once a cluster of five 
neighboring nodes with the same internal state has been established, this is stable over time. 
The question of con-or dissensus is thus equivalent to asking whether clusters with different 
states can appear (i.e. internal coalitions are built) which in turn crucially depends on the update 
sequence. Groups having passed the coefficient of inefficiency, as it is the case for 
$N=25$ and $N=35$, allow the formation of four clusters.
}
 \label{DStraj}
\end{figure}

These findings are closely related to the changes the topology undergoes with different group-sizes. 
For $N<10$ the network is fully connected - each member can directly influence each other. 
The choice of the update threshold $h$ assures that consensus can be reached in this range for 
each $N$ (with the exception of $N=8$). As the group grows there appear nodes which are not directly linked. 
Order phenomena emerge. A necessary prerequisite for dissensus is that there is a minority of at 
least five members (for the chosen $h$ and $k$, since $5>hk=0.6$). When five adjacent nodes hold the same 
state none of them can be updated 
anymore (since each of them will have maximally four neighbours in a different state which is 
not enough to reach a majority). In case one state is dominating in a local neighbourhood this 
may establish a stable cluster of at least five nodes, depending on the actual update sequence. 
These sensitivities concerning the initial distribution of nodes interplaying with the random 
sequence of updates makes it impossible to solve the model analytically, 
but on the other hand give rise to the observed nontrivial behaviour. 
With the avenue of a new group member more possibilities are opened up to establish stable 
clusters of different opinions. This is nothing but the forming of internal coalitions. 
It is straight-forward to see what happens if group size passes the critical region between 19 and 21. 
At this point two nodes arise which do not have any neighbours in common. 
Beyond this size also four internal groupings can be established. In other words, the number of 
ways to reach a dissensus has significantly expanded. The admission of one more member will 
thus have a lesser impact than in the smaller group. 
This constitutes the existence of a critical size which arises at the point where independent 
conversations between nodes can take place in the network. 
For large group sizes, as the maximal distance between two nodes increases (their correlation decreases), 
it becomes almost inevitable that balanced initial distributions lead to internal coalitions.
 These results hold, in principle, for every model of the opinion formation process which 
allows the formation of stable clusters, i.e. introduces (realistic) spatial correlations. 
Here this feature is incorporated by highly clustered small-world structure in combination with 
the random sequential updates.

The issue of group-size in decision-making processes is currently of imminent 
importance in the European Union, which has been subject to an enlargement 
to a club of now 27 nations. This is reflected in the executive branch of the 
union, the European Commission, numbering 27 members. This growth forced the union 
to reconsider its constitutional framework resulting in the Treaty of Lisbon, also known 
as Reform Treaty, where it is decided to reduce the Commission by 
two-thirds resulting in a group of 18 which would  bring the Commission (hopefully) below 
the coefficient of inefficiency.

The time has arrived to mention the case of $N=8$. As stated above this is the only feasible  
cabinet size which has been avoided by all countries now and fifty years ago. 
Without claiming any scientific relevance of this point, it is amusing that with our choice 
of $h=0.6$ is is possible to reproduce exactly this effect. In this case each node has seven 
neighbours and the network is fully connected. When the initial distribution is given by 
$A_i=4$ the majority seen by the members is $4/7 \approx 0.57 < h$, so the threshold is not exceeded. 
This accounts for one out of nine initial distributions and we find 
$D \left( 8 \right)=\textstyle \frac{1} {9} = 0.\overline{1}$. 
In British history this number was chosen only {\em once} for a cabinet \cite{Park57}. 
It might not come as a surprise that this occurred under the reign of Charles I, King of England, 
Scotland and Ireland, who became famous for being beheaded after advocating the 
Divine Right of Kings, levying taxes without the Parliament's consent and therefore triggering the First English Civil War \cite{Carlton95} .




\clearpage
 
\begin{figure*}[tb]
 \begin{center}
 \includegraphics[height=60mm]{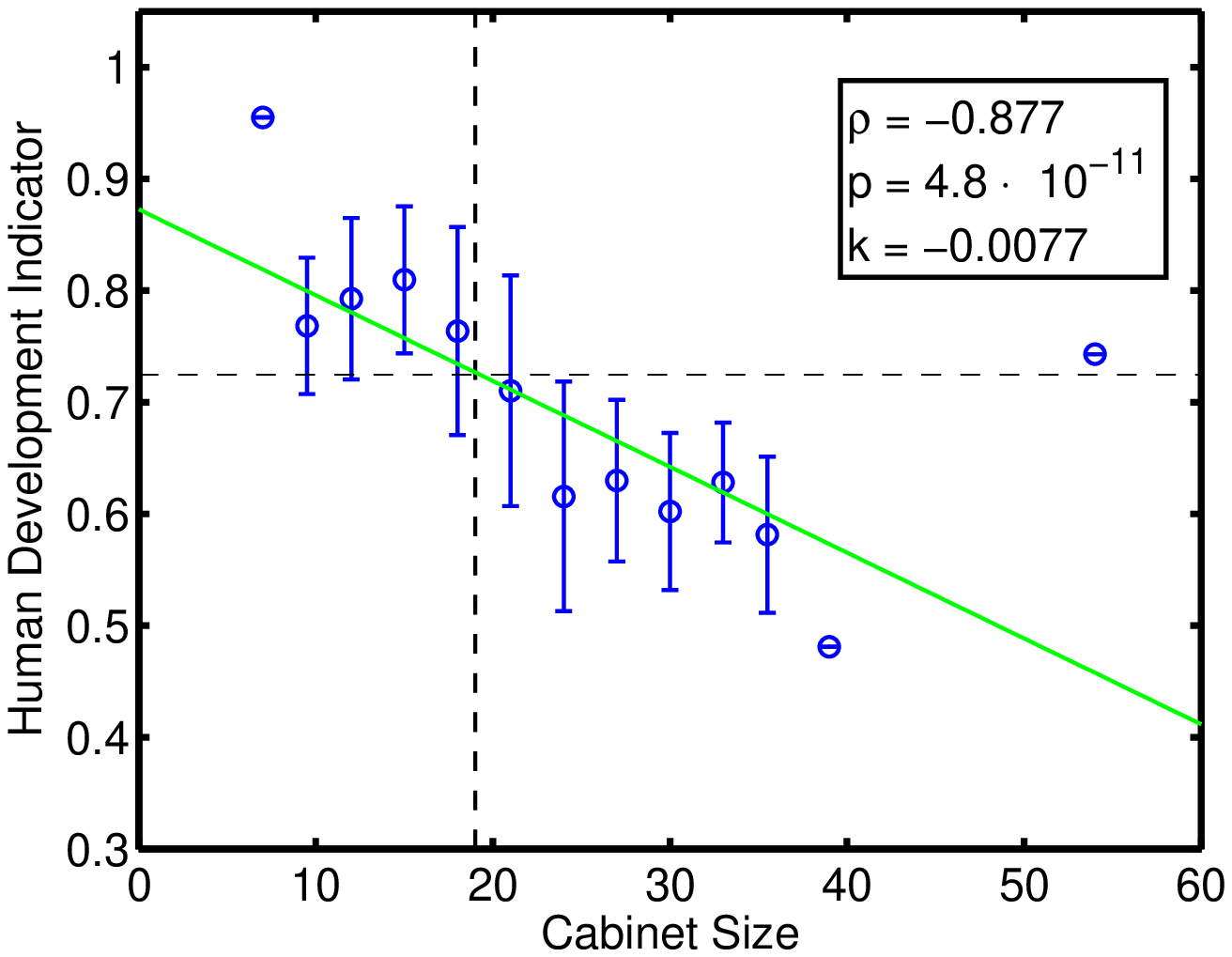}
 \includegraphics[height=60mm]{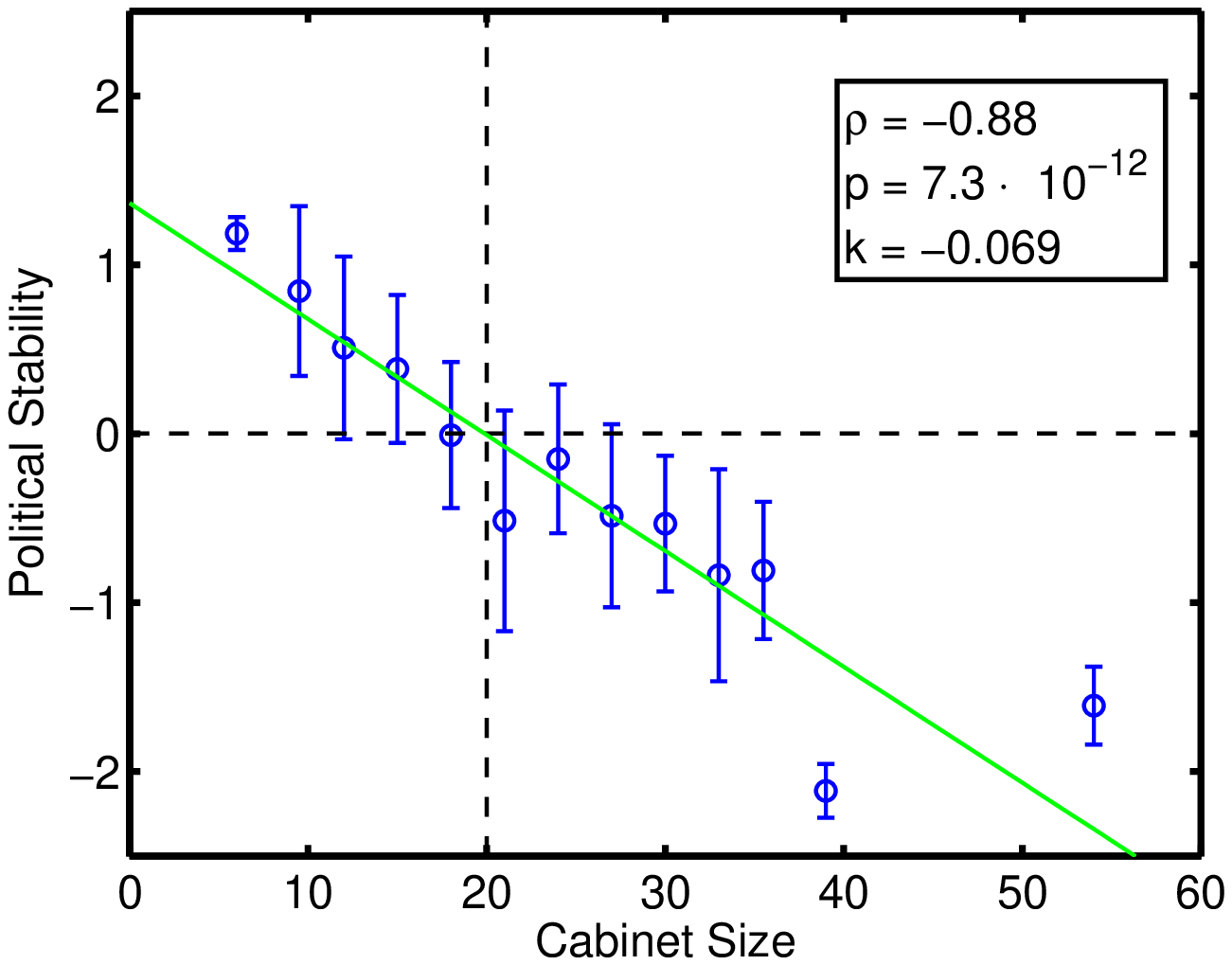}     \\
 \includegraphics[height=60mm]{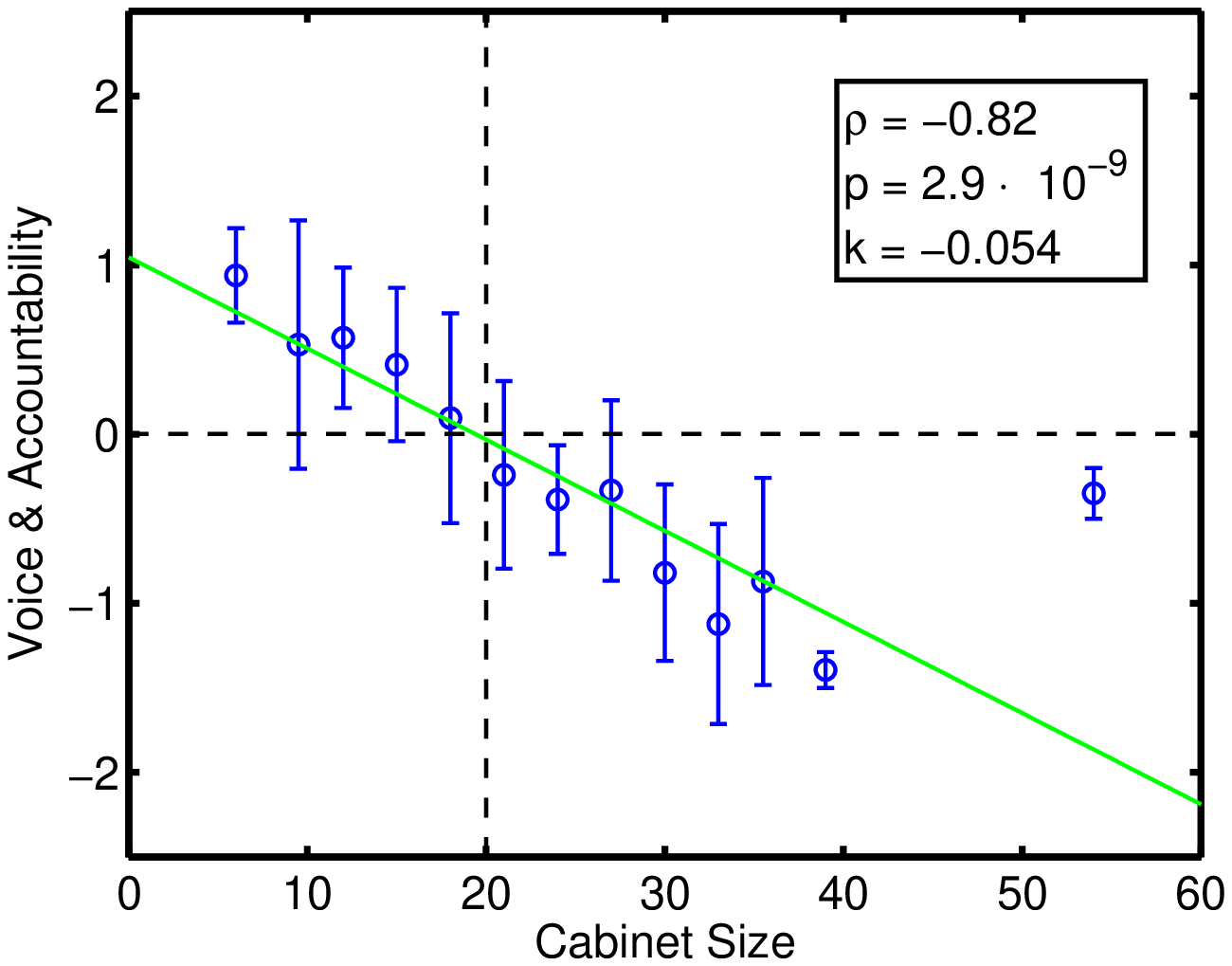}
 \includegraphics[height=60mm]{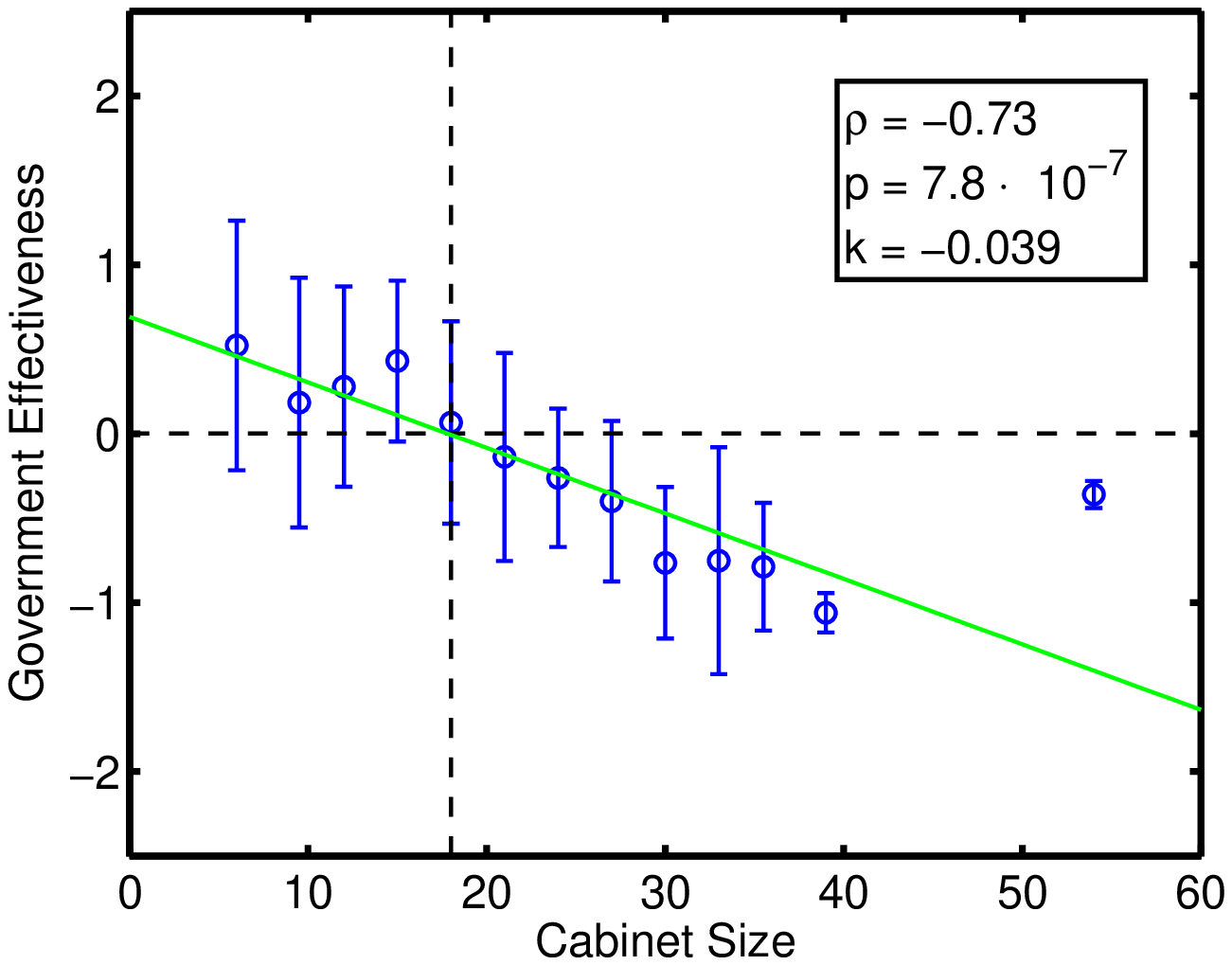}
 \end{center}
 \caption{The diagrams for the cabinet size versus the Human Development Indicator and the three governance indicators Political Stability, Voice \& Accountability and Government Effectiveness exhibit the same dependencies. For each indicator the correlation coefficient $\rho$, $p$-value and the slope $k$ are listed. The confidence levels vary between $p \leq 10^{-6}$ and $p \leq 10^{-11}$. The error bars show the standard deviations stemming from the averaging (see text), when the data comes from only one country the literature's standard deviation is used (with the exception of the HDI, where no error margins are provided). The horizontal dashed lines show the global average, the positions of the vertical lines are given by the intersections between the linear fit and this average. For all indicators these cabinet sizes are found between 18 and 20.}
 \label{Ierror}
\end{figure*}
\section{Appendix: Data and Methods}

\subsection{Cabinet Size}

The cabinet size for 197 self-governing countries and territories is extracted from a weekly updated database provided by the CIA \cite{chiefs} as of November 9, 2007. We are interested in the number of persons in the highest executive committee. This accounts for a country's cabinet where we counted the number of Minister or Secretaries including the Prime Minister (if he is a member of the cabinet, as it is mostly the case but not always, e.g. Switzerland)  and his vice(s). We do not include members of cabinets who hold a redundant office (e.g. Minister-Assistants or Minister of States). Attention has to be paid to the fact that in many cases the same person holds more than one office in a cabinet, we always count the number of persons and not offices. The obtained values are listed in table \ref{tabcab}. The only country where data is available but not included in our considerations is Bhutan. Here all but three members of the cabinet withdrew their office due to a new law stating the illegality of political party affiliation for cabinet members. The current caretaker regime does not formulate new governmental policies and only maintains day-to-day business.

\subsection{Human Development Indicator}

The Human Development Indicator (HDI) is published in the Human Development Report \cite{hdr} on behalf of the United Nations Development Programme (UNDP) on an annual basis. It compares the achievements in human development of 173 countries along three dimensions. The indicator is equally weighted composed of the standard of living (measured by the gross domestic product), knowledge (as given by the adult literacy rate and gross enrolment ratio) and a long and healthy life (given by the life expectancy at birth). Each index is normalized on a scale between 0 and 1, the HDI is then the arithmetical mean of those three. Unfortunately, no standard deviations are available.

\subsection{Governance Indicators}

The World Bank publishes annually six dimensions of governance in the Worldwide Governance Indicator research project. We use current data \cite{Kaufmann07} based on several hundred individual variables from 33 separate data sources by 30 different organisations. From this six aggregate indicators are constructed. We consider in the present work three dimensions of governance.

\emph{Political Stability and Absence of Violence.} This measures the perceptions of the likelihood that the government will be destabilised or overthrown by non-constitutional means. Aggregates for this indicator include the military coup risk, armed conflicts, social unrest, internal and external conflicts, government stability, political troubles, fractionalisation of the political spectrum, risk of political instability, etc.

\emph{Voice \& Accountability.} This measures to which extent people are able to participate in the selection of their government as well as basic human freedoms. Aggregates include political rights, freedom of the press, government censorship, military in politics, democratic accountability, institutional permanence, representativeness, hardening of the regime, transparency of government policies, etc.

\emph{Government Effectiveness.} This measures the quality of public and civil services, the degree of independence from political pressures and the quality of policy formulation. Aggregates are government instability and ineffectiveness, institutional failure, e-government, quality of bureaucracy, public spending composition, satisfaction with public transport systems, policy consistency and forward planning, management of public debt, health services and education, trust in government, etc. 
 
The governance indicators are measured in units following a normal distribution with zero mean and a standard deviation of one in each period, the vast majority of points lies between $-2.5$ and $2.5$, standard deviations are provided.

\begin{figure*}[thb]
 \begin{center}
 \includegraphics[height=60mm]{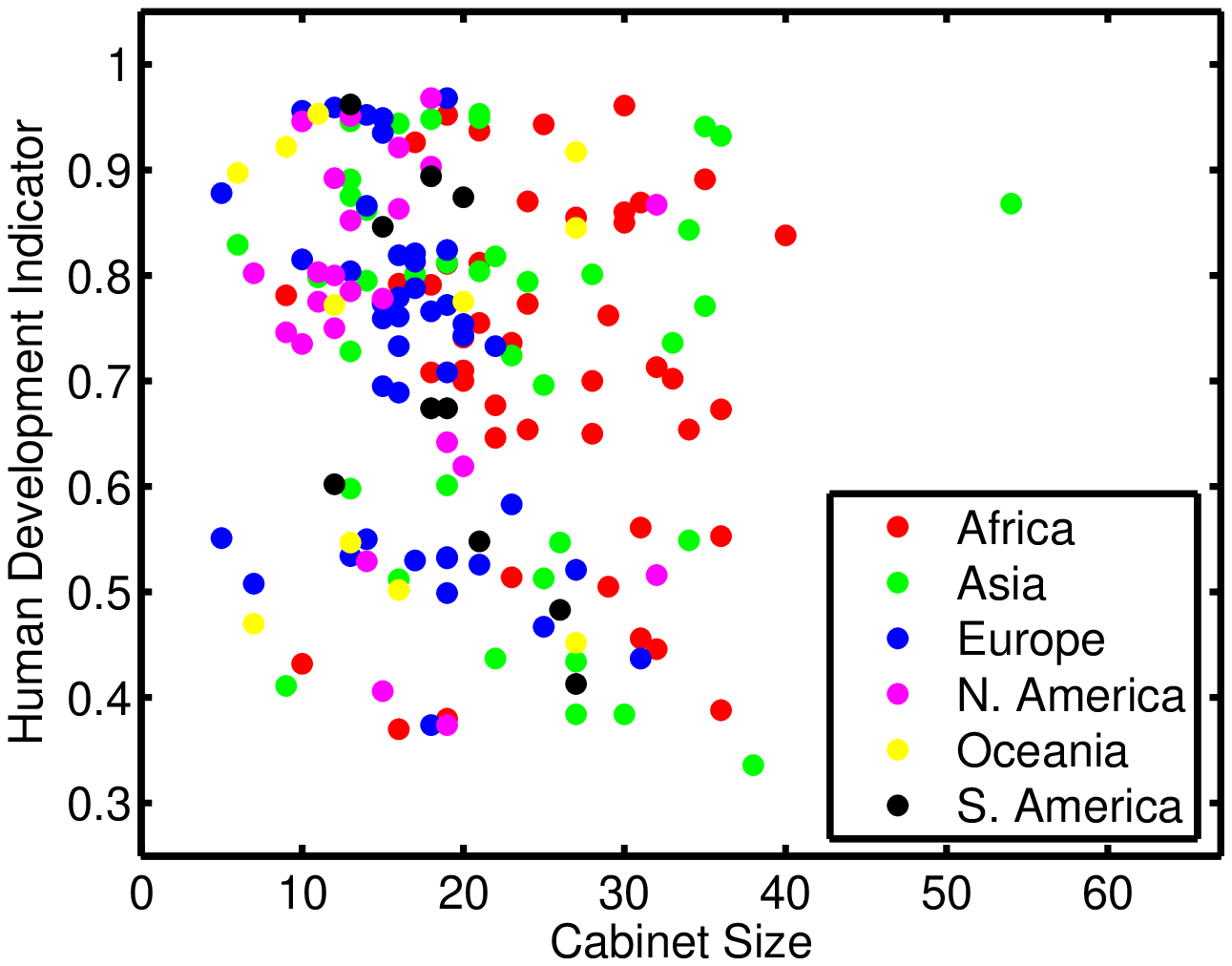}
 \includegraphics[height=60mm]{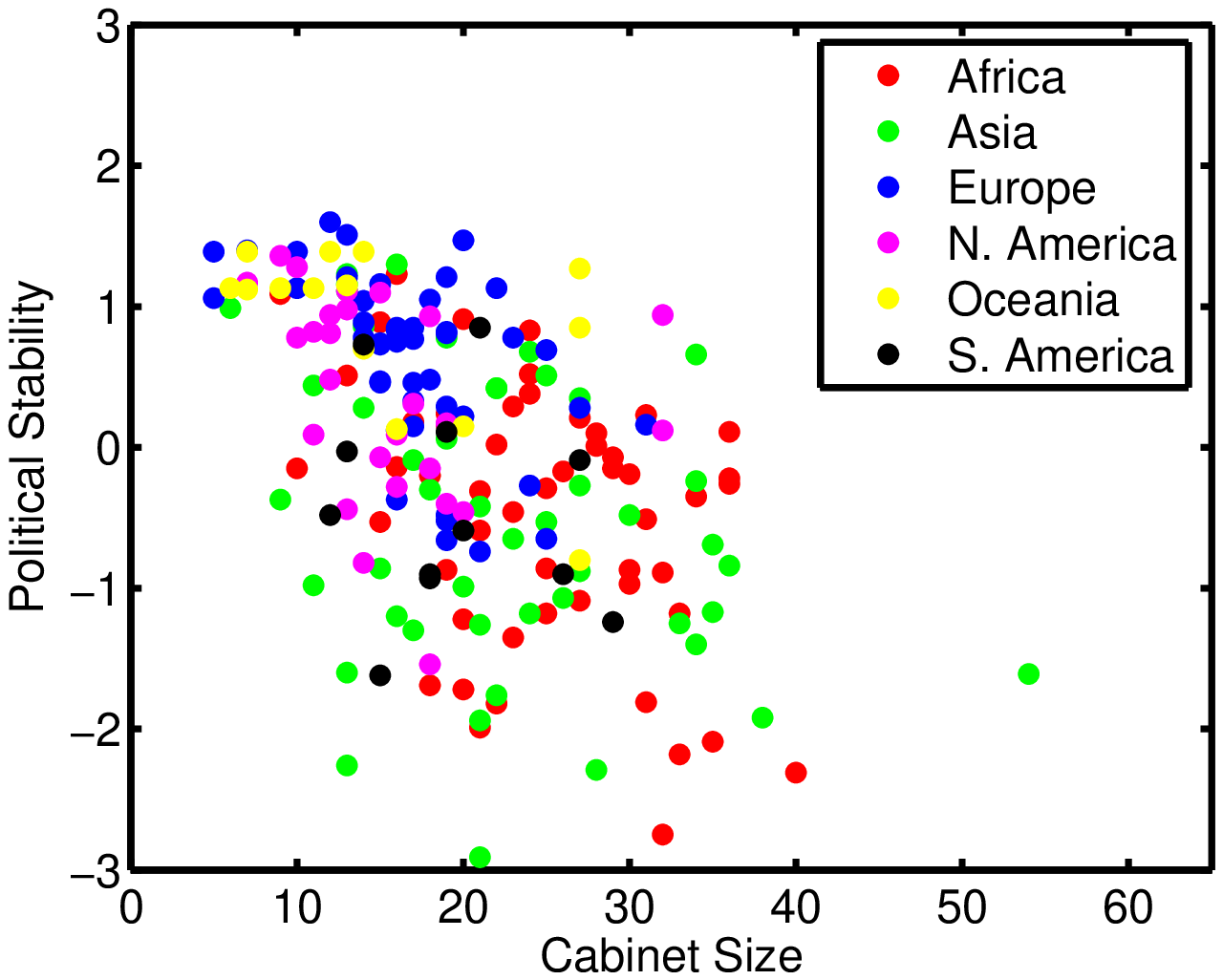}     \\
 \includegraphics[height=60mm]{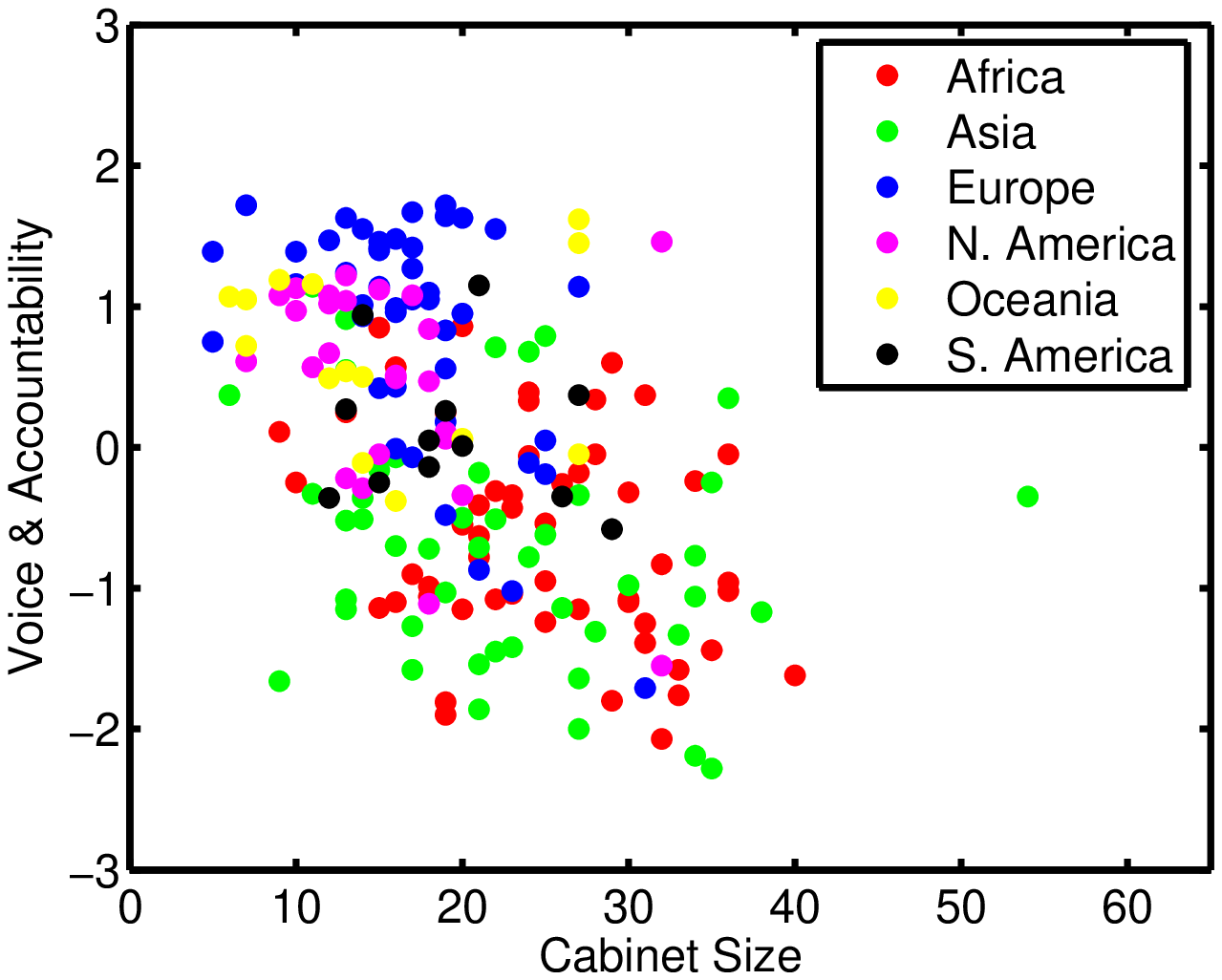}
 \includegraphics[height=60mm]{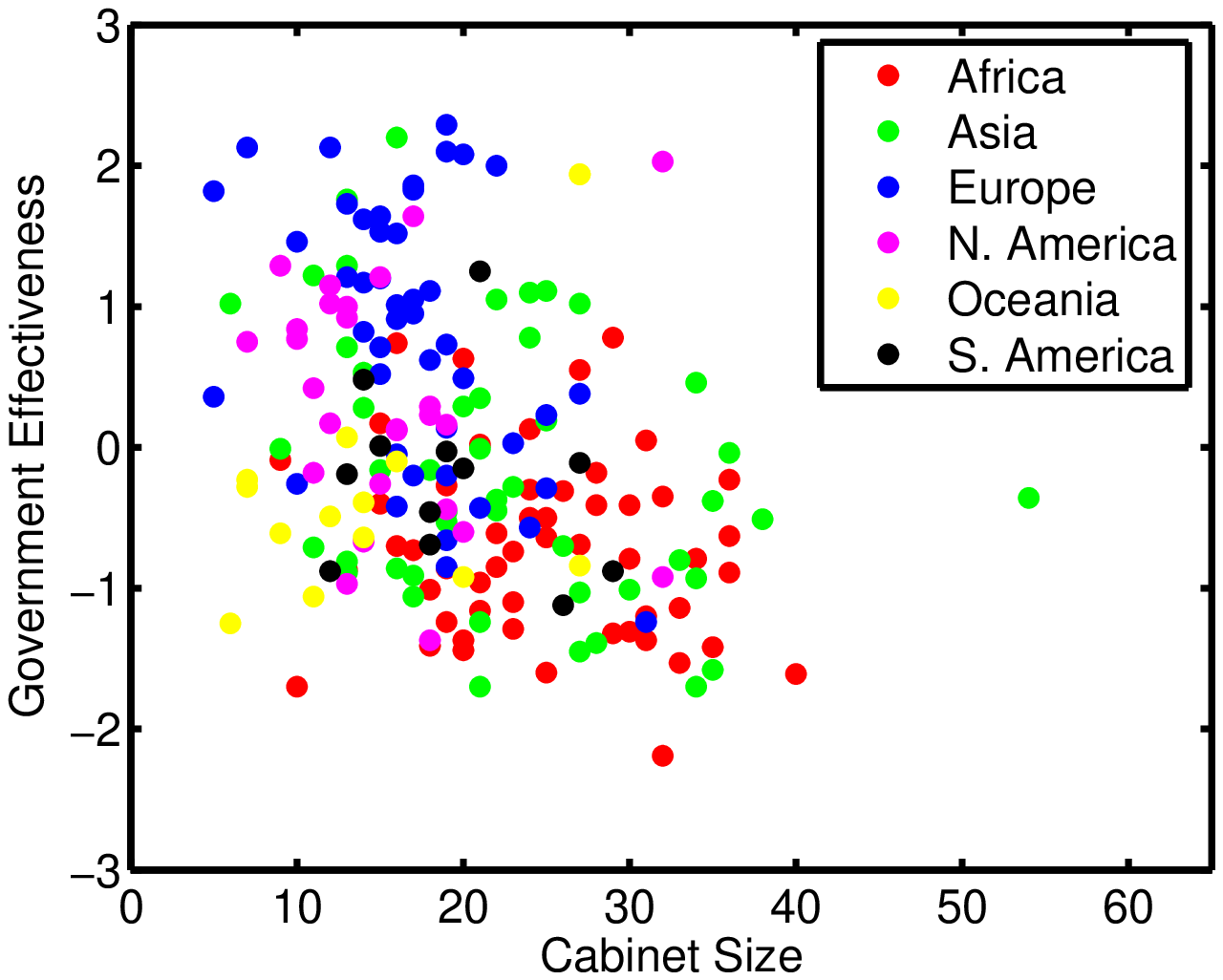}
 \end{center}
 \caption{The regional correlations in both, the indicators and the cabinet size, can also be seen from the raw data. Each point represents one country with coordinates given by the cabinet size and the respective indicator. The colour corresponds to the continent where the capital is found to be. Countries from Europe, America and Oceania dominate the north-western regions of the diagrams, countries from Asia and Africa are more likely to be found in the south-eastern regions.}
 \label{Iscatter}
\end{figure*}

\begin{figure}[tb]
 \begin{center}
 \includegraphics[height=55mm]{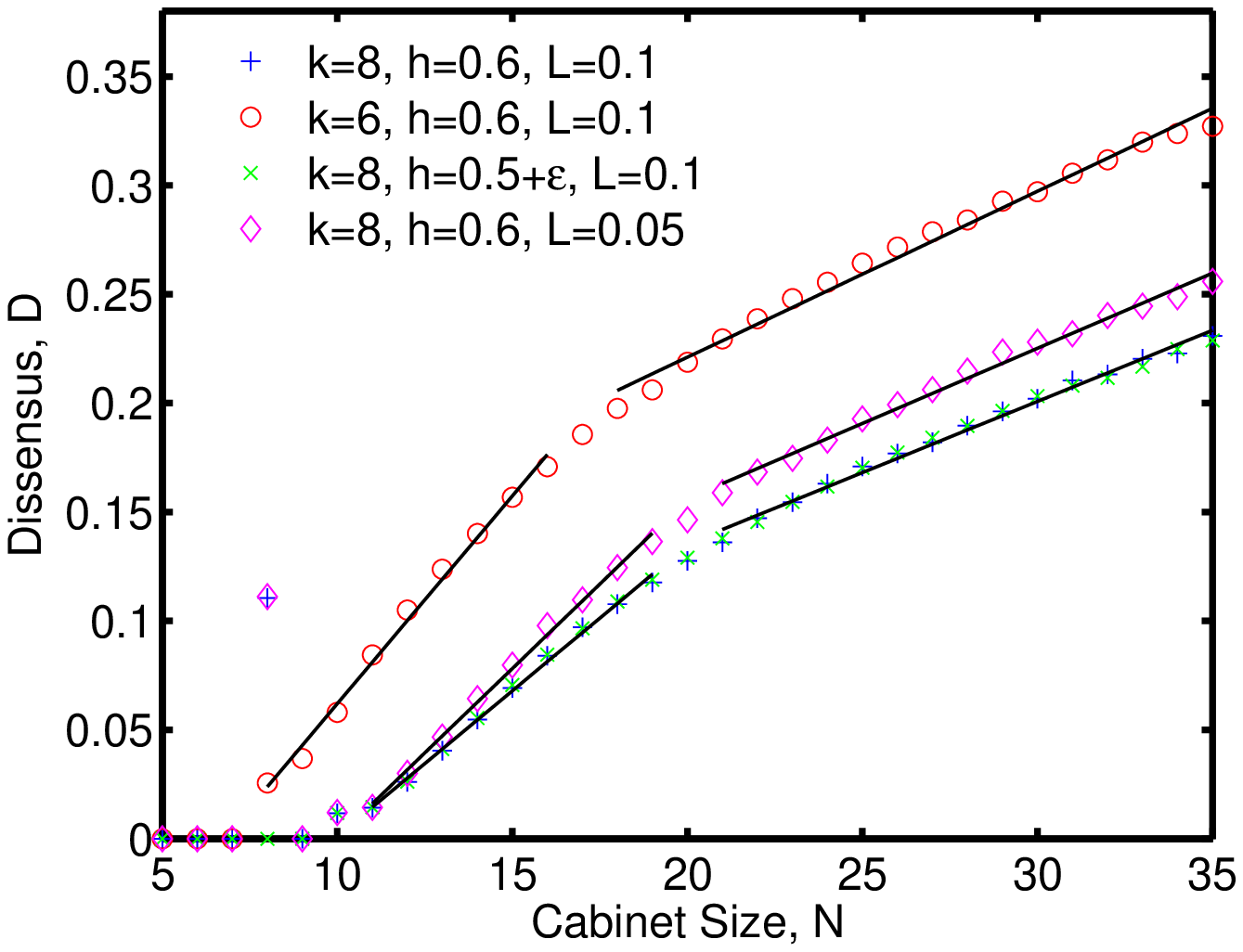}
 \end{center}
 \caption{$D\left(N\right)$ for four choices of the parameters $\left(k,h,L\right)$ is shown. As a reference we show $D\left(N\right)$ for $\left(8,0.6,0.1\right)$ again (pluses). Decreasing the connectivity of the group`s network, setting $\left(6, 0.6,0.1\right)$, circles, shifts the position of the critical point and increases the tendency toward dissensus. Using the pure majority rule, $\left(8,0.5+\epsilon,0.1\right)$, crosses, has no impact on $D\left(N\right)$ due to our choice of $k$, except that we do not find the 'Charles I' scenario. Lowering $L$ and therefore increasing the spatial correlations in the network hardens the finding of consensus too, settings $\left(8,0.6,0.05\right)$ (diamonds).}
 \label{DSpara}
\end{figure}

Figure \ref{Ierror} shows the interdependency between cabinet size and these indicators. For each size the mean values and standard deviations of the indicators are computed (standard deviations from the literature are used if only one country has the respective size, since this error is comparably small to the deviations from our averaging). From this values we computed the correlation coefficient $\rho$ and the $p$-value. Subsequently we bin the cabinet sizes with an interval of three and compute the error bars by Gaussian error propagation.

Let us discuss the implications of these correlations in more detail. These governance indicators give us a tool to investigate a country's political climate in more concise directions than the Human Development Indicator. The indicator Political Stability can be interpreted as a measure for the influence of constitutional and non-constitutional forces on a destabilisation of the government and is therefore related to the number of interests and interest groups that have to be satisfied. This is also reflected in the composition of the cabinet, thus the negative correlation with the cabinet-size. A naive interpretation of our results makes the conclusion tempting that a dictatorship would be the most effective form of decision-making. The indicator Voice \& Accountability, however, reveals that exactly the opposite holds. It quantifies to which extent citizens have elected their current leaders. A country with a low value here is thus more likely to be reigned by a sovereign leader or council which often confers executive, legislative and jurisdicative powers onto one hand. In this case the nominal executive council, the cabinet, is less influent and important than in countries where it is indeed the highest executive council. We find that this tends to increase the membership of the cabinet which can be understood through the minor importance and therefore exclusivity of it. Government Effectiveness gives us insight into the quality of policy formulation in the government and is thus directly related to a cabinet's ability to find consensus on an issue in question and advance it reasonably. Let us stress that the actual size of the cabinet is not included in the aggregates. Furthermore, this indicator also measures how efficient this policies are implemented from the government downward to the citizens.  

In figure \ref{Iscatter} we show the raw data for the indicators versus cabinet size. A colour code for the continents shows regional clustering of the points. Countries from Europe, America and Oceania are more likely to be found in the north-western region of the plots than countries from Asia and Africa.

\section{Simulation Details}

The numerical results are obtained by counting the frequency of final configurations without consensus out of $10^5$ realisations. In each run we first create a regular 1D ring where each node is connected to its $k$ nearest neighbors. Each link is then deleted with probability $L$ and new links are randomly created such that there are $Nk$ links in total and each node has exactly $k$ links again. 

\subsection{Influence of the model parameter.}
The number of neighbours $k$ determines the position of the critical point.
The driving mechanism is the allowance for internal coalitions, i.e. the formation of stable clusters. When the network is fully connected we either encounter consensus or a frozen system, depending on $h$. Our choice of $h=0.6$ is primarily motivated by giving rise to a frozen state for a fully connected network with $N=8$ and a balanced initial distribution, i.e. an equal number of nodes being initially in state 0 and 1. For other choices one may encounter different frozen states. For fixed model parameters and increasing group sizes there is always a point where stable clusters begin to emerge. It is this point where the increase in dissensus not stemming from an initially frozen state sets in. Note that for our choices in the case of $N=10$ the main contributions in dissensus still com from frozen systems, here an evolution toward a dissensus state is highly unlikely. Finally the critical point, where an increase in group size leads to considerably smaller increments in the dissensus, can be found when four cluster can be formed for given $h$ and $k$.

With increasing $L$ the topology becomes less regular and approaches a random graph for $L=1$. In other words, the neighbourhoods of two neighbouring nodes are becoming more independent with increasing $L$ and local correlations diminish. Consensus can be easier reached for networks with higher $L$. Figure \ref{DSpara} shows $D\left(N\right)$ for different parameter settings and confirms the above stated observations. We show four different settings of the model parameters $\left(k,h,L\right)$. As a reference we show $D\left(N\right)$ for $\left(8,0.6,0.1\right)$ again (pluses). Decreasing the connectivity of the group`s network, parameter setting $\left(6, 0.6,0.1\right)$, circles, shifts the position of the critical point and increases the tendency toward dissensus. Adjusting the threshold such that we recover the pure majority rule, $\left(8,0.5+\epsilon,0.1\right)$, crosses, has no impact on $D\left(N\right)$ due to our choice of $k$, except that we do not find the 'Charles I' scenario in this case. Of course, for the same $k,L$ and $h>5/8=0.6125$ we would find an increase in dissensus. Lowering $L$ and therefore increasing the spatial correlations in the network hardens the finding of consensus too, as can be seen from the settings $\left(8,0.6,0.05\right)$ (diamonds).




\begin{table*}[p]
\begin{tabular} {l | p{140mm}}
Members & Countries \\
\hline
5 & Liechtenstein, Monaco  \\
\hline
6 & Macao, Nauru \\
\hline
7 & Cook Islands, Micronesia, Netherlands Antilles, Switzerland, Tuvalu \\
\hline
8 & - \\
\hline
9 & Aruba, China, Palau, Seychelles \\
\hline
10 & Andorra, Comoros, Dominica, Saint Kitts and Nevis, San Marino \\
\hline
11 & Antigua and Barbuda, Belize, Cyprus, Marshall Islands, Timor-Leste \\
\hline
12 & Bahamas, Bermuda, Grenada, Iceland, Kiribati, Paraguay, Saint Vincent and Grenadines \\
\hline
13 & Argentina, Bangladesh, Brunei, Hong Kong, Japan, Luxembourg, Malta, Nepal, Nicaragua, Saint Lucia, Sao Tome and Principe, Samoa \\
\hline
14 & Austria, Estonia, Guatemala, Kuwait, Lithuania, Quatar, Tonga, Uruguay, Vanatu \\
 \hline
15 & Barbados, Belgium, Cape Verde, Colombia, Croatia, El Salvador, France, Georgia, Hungary, Ireland, Rwanda \\
\hline
16 & Albania, Botswana, Czech Rep., Fiji, Germany, Jamaica, Kyrgyzstan, Panama, Romania, Singapore, Slovakia, Swaziland  \\
\hline
17 & Gambia, Laos, Montenegro, Netherlands, Portugal, Spain, Tajikistan, United Kingdom, United States \\
\hline
18 & Armenia, Bolivia, Central African Rep., Costa Rica, Djibouti, Greece, Haiti, Peru, Slovenia, Trinidad and Tobago \\
\hline
19 & Bosnia and Herzegovina, Bulgaria, Denmark, Dominican Rep., Eritrea, Kazakhstan, Latvia, Lesotho, Libya, Macedonia, Mexico, Moldova, Mongolia, Norway, Suriname \\
\hline
20 & Finland, Guinea, Guyana, Honduras, Liberia, Mauritius, Poland, Solomon Islands, Thailand \\
\hline
21 & Bahrain, Chile, Guinea-Bissau, Iraq, Morocco, Nigeria, Philippines, Russia, Uzbekistan \\
\hline
22 & Ethiopia, Korea (South), Lebanon, Malawi, Sweden, Vietnam \\
\hline
23 & Burundi, Maldives, Saudi Arabia, Sierra Leone, Zambia \\
\hline
24 & Benin, Israel, Mozambique, Namibia, Ukraine, United Arabian Emirates \\
\hline
25 & Jordan, Mauritania, Serbia, Taiwan, Togo, Turkey, Uganda \\
\hline
26 & Azerbaijan, Ecuador, Tanzania \\
\hline
27 & Australia, Brazil, Italy, Kenya, Malaysia, New Zealand, Papua New Guinea, Syria, Tunisia, Turkmenistan \\
\hline
28 & Afghanistan, Madagascar, Mali \\
\hline
29 & Equatorial Guinea, South Africa, Venezuela \\
\hline
30 & Burkina Faso, Cambodia, Congo (Rep. of), Egypt \\
\hline
31 & Angola, Belarus, Chad, Ghana \\
\hline
32 & Algeria, Canada, Cuba, Somalia \\
\hline
33 & Iran, Sudan, Zimbabwe \\
\hline
34 & Korea (North), Niger, Oman, Yemen \\
\hline
35 & Burma (Myanmar), Cote d'Ivoire, Indonesia \\
\hline
36 & Cameroon, Gabon, India, Senegal \\
\hline
38 & Pakistan \\
\hline
40 & Congo (Dem. Rep. of) \\
\hline
54 & Sri Lanka \\
\hline
\end{tabular}
\caption{Number of members on the highest level of the highest executive committee \cite{chiefs}, harvested at 10/09/2007.}
\label{tabcab}
\end{table*}

\end{document}